\documentclass[preprint,pra,showpacs,nofootinbib]{revtex4}
\usepackage[mathscr]{euscript}
\usepackage{graphicx}
\usepackage{float,epsfig}
\usepackage{dcolumn}% Align table columns on decimal point
\usepackage{bm}% bold math
\usepackage{graphicx}% Include figure files
\usepackage{dcolumn}% Align table columns on decimal point
\usepackage{amsmath,amssymb,amsthm}
\usepackage[colorlinks=true,linkcolor=blue]{hyperref}
\usepackage{booktabs}

\newcommand{\bea}{\begin{eqnarray}}
\newcommand{\eea}{\end{eqnarray}}
\newcommand{\beq}{\begin{equation}}
\newcommand{\eeq}{\end{equation}}

\def\/{\over}
\begin{document}
\title{Geometric phase outside a Schwarzschild black hole and the Hawking effect}
\author{  Jiawei Hu$^{1}$ and Hongwei Yu$^{1,2}$\footnote{Corresponding author} }
\affiliation{$^1$ Institute of Physics and Key Laboratory of Low Dimensional Quantum Structures and Quantum
Control of Ministry of Education,\\
Hunan Normal University, Changsha, Hunan 410081, China \\
$^2$ Center for Nonlinear Science and Department of Physics, Ningbo University, Ningbo, Zhejiang 315211, China}

%\date{\today}

\begin{abstract}

We study the Hawking effect in terms of the geometric phase acquired by a two-level atom as a result
of coupling to vacuum fluctuations outside a Schwarzschild black
hole in a gedanken experiment. We treat the atom in interaction with a bath of fluctuating
quantized massless scalar fields as an open quantum system, whose
dynamics is governed by a master equation obtained by tracing over
the field degrees of freedom. The nonunitary effects of this system
are examined by analyzing the geometric phase  for the Boulware,
Unruh and Hartle-Hawking vacua respectively. We find, for all the
three cases, that the geometric phase of the atom turns out to be
affected by the space-time curvature which  backscatters  the vacuum
field modes. In both the Unruh and Hartle-Hawking vacua, the
geometric phase exhibits similar behaviors as if there were thermal
radiation at the Hawking temperature from the black hole. So,  a
measurement of the change of the geometric phase as opposed to that
in a flat space-time can in principle reveal the existence of the
Hawking radiation.

\end{abstract}

\maketitle

\section{Introduction}

In quantum sense, a black hole is not completely black,  but emits
thermal radiation with a black body spectrum.  This is the
well-known Hawking radiation which is first found by Hawking in the
paradigm of quantum field theory in curved space-time~\cite{hk}.
Ever since Hawking's original work, this striking effect has
attracted a great deal of interest, and extensive work has been done
trying to re-derive and understand it in various different
contexts~\cite{hk,Gibbons,Parikh,SC,Robinson,hkur,yu1,yu2,yu3,hu1}, ranging from Euclidean quantum gravity~\cite{Gibbons} to quantum entanglement~\cite{hu1}.

In this paper, we will try to understand, in the framework of open quantum systems,  the Hawking effect by
studying the geometric phase acquired by a static two-level atom due
to its coupling to the vacuum fluctuations outside a Schwarzschild
black hole in a gedanken experiment. The geometric
phase is an important concept in quantum theory. In 1984, Berry
showed that when the Hamiltonian of a closed quantum system evolves
adiabatically in a cyclic way, the state of the system acquires an
additional phase besides the familiar dynamical one, which has a
purely geometric origin~\cite{berry}.  Berry's work was soon
generalized to nonadiabatic~\cite{nonadiabatic} and noncyclic
evolution~\cite{noncyclic}.  The geometric phase has been
extensively studied both theoretically and
experimentally~\cite{GPbook}. In recent years, it has been
demonstrated that geometric phase may have many potential
applications, for instance, fault-tolerant quantum
computation~\cite{fault-tolerant}. However, due to the inevitable
interactions between the qubits and the environment, a pure state
will be driven to a mixed one. Therefore, the study of geometric
phase of systems under nonunitary evolutions becomes important, and
some work has been done to generalize the geometric phase to open
quantum systems~\cite{Uhlmann, Sjoqvist, Singh, Tong, Wang}, and the
effects of different kinds of decoherence sources on the geometric
phase have been analyzed~\cite{Carollo, Rezakhani, Lombardo, chen,
Marzlin}.  By studying the geometric phase of an open system, we can
draw some information about the nonunitary evolution of the system
and the characteristics of the environment the system coupled
to.  In principle, the geometric phase factor can  be measured by
interfering the system which has undergone the above evolution with
one that does not. Noteworthily,  a new
type of experiment on the quantum geometric phase of an open system
undergoing nonunitary evolution has recently been done, where the
phase is determined by measuring the decoherence factor of the
off-diagonal elements of the reduced density matrix of the
system~\cite{critical}.

Recently, Martin-Martinez et al.~\cite{Martin} have studied the
geometric phase of an accelerated atom which couples to a
single-mode of a scalar field in vacuum and found that the phase
differs from that of an inertial one as a direct consequence of the
Unruh effect. Thus, in principle, an interferometry experiment can
be performed to reveal the effects of acceleration. Later, we have
considered a more realistic case, i.e., an accelerated two-level
system which couples to all vacuum modes of electromagnetic (rather
than scalar ) fields in a realistic multipolar coupling scheme,  and
proposed using the geometric phase of non-unitary evolution to detect
the Unruh effect~\cite{hu2}. In the current paper, we study the
geometric phase acquired by a two-level atom interacting with a bath of
fluctuating quantized massless scalar fields in vacuum outside a
Schwarzschild black hole in a gedanken experiment, in the hope of gaining understanding of
the Hawking effect in a new perspective.  In the gedanken experiment,  we first prepare the atom in a pure state,  adiabatically put it at a fixed radial
distance outside the black hole, then the atom interacts
with the fluctuating scalar fields in vacuum, and as a result, it evolves from a
pure state to a mixed state. During this nonunitary evolution, a
geometric phase is acquired. The atom in the gedanken experiment can be thought of  as being held static by an imaginary string with tension, or the the
buoyancy of the Hawking radiation, or even some other unknown mechanism.
 The reduced dynamics of the
atom is studied by tracing over the field degrees of freedom from
the total system. The Boulware, Unruh and Hartle-Hawking vacua will
be considered. At this point, let us note that the theory of open quantum
system has been fruitfully applied to understand the
Unruh~\cite{Benatti1}, Hawking~\cite{yu3}, and
Gibbons-Hawking~\cite{yu4} effects respectively in a way that is
different from the traditional one.

\section{the master equation}

The Hamiltonian of the whole system takes the form
\beq
H=H_s+H_\phi+H'\;.
\eeq
Here $H_s$ is the Hamiltonian of the atom, and its general form is taken to be
\beq
H_s={\omega_0\over 2}\sum_{i=1}^3\,n_i\,\sigma_i\;,
\eeq
in which $\sigma_i$ is the Pauli matrix, and $\omega_0$ is the energy level spacing of the atom. For simplicity, we can write it as $H_s={1\over 2}\omega_0\sigma_3$. $H_\phi$ is the Hamiltonian of the scalar field, whose explicit expression is not required here. We couple the atom and the scalar field by analogy with the electric dipole interaction~\cite{Hint}
\begin{equation}
H'=\mu\,\sigma_2\,\Phi(t,{\bf x})\;,
\end{equation}
where $\mu$ is the coupling constant, which we assume to be small.

The initial state of the whole system is characterized by the total density matrix $\rho_{tot}=\rho(0) \otimes |0\rangle\langle0|$, in which $\rho(0)$ is the initial reduced density matrix of the atom, and $|0\rangle$ is the vacuum state of the field. In the frame of the atom, the evolution in the proper time $\tau$ of the total density matrix $\rho_{tot}$ satisfies
\begin{equation}
\frac{\partial\rho_{tot}(\tau)}{\partial\tau}=-{i\/\hbar}[H,\rho_{tot}(\tau)]\;.
\end{equation}
 The evolution of the reduced density
matrix $\rho(\tau)$ can, in the limit of weak coupling,  be written
in the Kossakowski-Lindblad form~\cite{Lindblad, Benatti1, Benatti2,
pr5}
\begin{equation}\label{master}
{\partial\rho(\tau)\over \partial \tau}= -{i\/\hbar}\big[H_{\rm eff},\,
\rho(\tau)\big]
 + {\cal L}[\rho(\tau)]\ ,
\end{equation}
where
\begin{equation}
{\cal L}[\rho]={1\over2} \sum_{i,j=1}^3
a_{ij}\big[2\,\sigma_j\rho\,\sigma_i-\sigma_i\sigma_j\, \rho
-\rho\,\sigma_i\sigma_j\big]\ .
\end{equation}
The matrix $a_{ij}$ and the effective Hamiltonian $H_{\rm eff}$ are determined by the Fourier and Hilbert transforms of the correlation functions
\begin{equation}\label{correlation}
{G^+}(x-y)=\mu^2\langle0|\Phi(x)\Phi(y)|0 \rangle\;,
\end{equation}
which are defined as follows
\begin{equation}
{\cal G}(\lambda)=\int_{-\infty}^{\infty} d\tau \,
e^{i{\lambda}\tau}\, {G^+}(\tau)\; , \label{fourierG}
\end{equation}
\begin{equation}
{\cal K}(\lambda)=
\frac{P}{\pi i}\int_{-\infty}^{\infty} d\omega\ \frac{ {\cal
G}(\omega) }{\omega-\lambda} \;,
\end{equation}
in which $P$ denotes principal value. Then the coefficients of the Kossakowski matrix $a_{ij}$ can be expressed as~\cite{Benatti1}
\begin{equation}
a_{ij}=A\;\delta_{ij}-iB\; \epsilon_{ijk}n_k+C\;n_in_j\; ,
\label{cij}
\end{equation}
where
\begin{equation}
A=\frac{1}{4}[{\cal {G}}(\omega_0)+{\cal{G}}(-\omega_0)]\;,\;~~~
B=\frac{1}{4}[{\cal {G}}(\omega_0)-{\cal{G}}(-\omega_0)]\;,~~~~
C=-A\;.\ \label{abc}
\end{equation}
The effective Hamiltonian $H_{\rm eff}$ contains a correction term, the so-called Lamb shift, and one can show that it can be obtained by replacing $\omega_0$ in $H_s$ with a renormalized energy level spacing $\Omega$ as follows~\cite{Benatti1}
\begin{equation}\label{heff}
H_{\rm eff}=\frac{1}{2}\hbar\Omega\sigma_3={\hbar\over 2}\{\omega_0+{i\/2}\,[{\cal
K}(-\omega_0)-{\cal K}(\omega_0)]\}\,\sigma_3\;.
\end{equation}

For convenience, the density matrix $\rho$ is expressed in terms of the Pauli matrices,
\begin{equation}\label{density}
\rho({\tau})=\frac{1}{2}\bigg(1+\sum_{i=1}^{3}\rho_i({\tau})\sigma_i\bigg)\;.
\end{equation}
Taking Eq.~(\ref{density}) into Eq.~(\ref{master}), one can show
that the components of the Bloch vector $|\rho(\tau)\rangle$ satisfy
\begin{equation}
\frac{\partial}{\partial{{\tau}}}|\rho({\tau})\rangle=-2{\cal{H}}|\rho({\tau})\rangle+|\eta\rangle\;,
\end{equation}
where $|\eta\rangle=\{0,0,-4B\}$, and ${\cal{H}}$ takes the form
\begin{equation}
{\cal{H}}=\left(
\begin{array}{ccc}
2A+C& \Omega/2& 0\\ -\Omega/2&2A+C&0 \\ 0&0&2A
\end{array}\right)\;.
\end{equation}
Now we assume that the initial state of the atom is
$|\psi(0)\rangle=\cos{\theta\/2}|+\rangle+\sin{\theta\/2}|-\rangle$, and
then it evolves from a pure state to a mixed state due to interaction with the external environment, namely, a bath of fluctuating
vacuum scalar fields.  The
time-dependent reduced density matrix of the atom can be easily
worked out as
\begin{equation}\label{dens}
\rho(\tau)=\left(
\begin{array}{ccc}
e^{-4A\tau}\cos^2{\theta\/2}+{B-A\/2A}(e^{-4A\tau}-1) & {1\/2}e^{-2(2A+C)\tau-i\Omega\tau}\sin\theta\\ {1\/2}e^{-2(2A+C)\tau+i\Omega\tau}\sin\theta & 1-e^{-4A\tau}\cos^2{\theta\/2}-{B-A\/2A}(e^{-4A\tau}-1)
\end{array}\right)\;.
\end{equation}

\section{geometric phase in open two-level system}

In this section, we use the master equation governing the reduced
dynamics of the atom in the last section to calculate  the geometric
phase of a static atom in the background of a Schwarzschild
space-time, which is described by the following metric
\begin{equation}
ds^2=\bigg(1-\frac{2M}{r}\bigg)dt^2-\frac{dr^2}{1-2M/r}-r^2(d\theta^2+\sin^2\theta{d\phi^2})\;.
\end{equation}
When  a curved space-time in concerned, a delicate issue is how to
determine the vacuum state of the quantum fields. Normally, a vacuum
state is associated with non-occupation of positive frequency modes.
However, the positive frequency of field modes is defined with
respect to the time coordinate. Therefore, to define positive
frequency, one has to first specify a definition of time. In a
spherically symmetric black hole background, three different vacuum
states, i.e., the Boulware~\cite{Boulware}, Unruh~\cite{Unruh}, and
Hartle-Hawking \cite{Hartle-Hawking} vacuum states, have been
defined, each corresponding to a different choice of time
coordinate. In the following, we examine the geometric phase of the
atoms in all these vacua.

\subsection{The Boulware vacuum}

Let us begin our discussion with the Boulware vacuum, which is
defined by requiring normal modes to be positive frequency with
respect to the Killing vector $\partial/\partial t$. In order to
analyze the geometric phase of the system, we make use of the master
equation Eq.~(\ref{master}) and see how the phase is related to it.

The geometric phase for a mixed state under nonunitary evolution is
defined as~\cite{Tong}
\beq\label{gp} \gamma=\arg \left(
\sum\limits_{k=1}^N \sqrt{\lambda_k(0)\lambda_k(T)}\langle
\phi_k(0)|\phi_k(T)\rangle e^{-\int_0^T \langle \phi_k(\tau)|\dot
\phi_k(\tau) \rangle d\tau} \right)\;, \eeq
 where $\lambda_k(\tau)$
and $|\phi_k(\tau)\rangle$ are the eigenvalues and eigenvectors of
the reduced density matrix $\rho(\tau)$. In order to get the
geometric phase, we first calculate the eigenvalues of the density
matrix (\ref{dens}) to find $\lambda_\pm(\tau)={1\/2}(1\pm\eta)\;, $
in which $\eta=\sqrt{\rho_3^2+e^{-4(2A+C)\tau}\sin^2\theta}$ and
$\rho_3=e^{-4A\tau}\cos\theta+{B\over A}(e^{-4A\tau}-1)$. It is
obvious that $\lambda_-(0)=0$.  As a result, the only contribution
comes from the eigenvector corresponding to $\lambda_+$ \beq
|\phi_+(\tau)\rangle=\sin{\theta_{\tau}\/2}|+\rangle+\cos{\theta_{\tau}\/2}e^{i\Omega\tau}|-\rangle\;,
\eeq where \beq\label{tan}
\tan{\theta_{\tau}\/2}=\sqrt{\eta+\rho_3\/\eta-\rho_3}\;. \eeq The
geometric phase can be calculated directly using Eq.~(\ref{gp})
\beq\label{gp1}
\gamma=-\Omega\int_0^T\cos^2{\theta_{\tau}\/2}\,d\tau\;. \eeq
Plugging the explicit form of $\theta_\tau$, $\eta$ and $\rho_3$
into Eq.~(\ref{gp1}), we have \beq\label{bd}
\gamma=-\int_0^{T}{1\over2}\bigg(1-\frac{R-R\,e^{4A\tau}+\cos\theta}{\sqrt{e^{-4C\tau}\sin^2\theta
+(R-R\,e^{4A\tau}+\cos\theta)^2}}\bigg)\,\Omega\,d\tau\;. \eeq
 The result of this integral is rather tedious. Here, for small
$\gamma_0/\omega_0$, we perform a series expansion of the geometric
phase for a single quasi-cycle and obtain to the first
order~\cite{hu2}
\beq\label{GPb} \gamma\approx-\pi(1-\cos\theta)+{2\pi^2\/\omega_0}
(2B-C\cos\theta)\sin^2\theta\;. \eeq
 It is obvious that the
geometric phase of the system is characterized by the coefficients
$B$ and $C$ which determine the Kossakowski matrix $a_{ij}$,  and we
now  calculate them explicitly.

Let us now calculate the geometric phase of a two-level atom which
is held static at a fixed radial distance by some additional force, an imaginary string with tension, for instance,
outside a Schwarzschild black hole. The trajectory can be described
by
 \beq\label{traj} t(\tau)={1\/\sqrt{g_{00}}}(\tau-\tau_0),\ \ \
r(\tau)=r_0,\ \ \ \theta(\tau)=\theta_0,\ \ \ \phi(\tau)=\phi_0\;.
 \eeq
In the Boulware vacuum, the Wightman function for the scalar field
on the trajectory Eq.~(\ref{traj}) is given
by~\cite{wightman1,wightman2,wightman3}
\begin{equation}
 G^+(x,x')=\mu^2\sum_{lm}\int_{0}^{\infty}\frac{e^{-i\omega \Delta{t}}}{4\pi\omega}|\,Y_{lm}(\theta,\phi)\,|^2\big[|\,\overrightarrow{R_l}(\omega,r)\,|^2
+|\,\overleftarrow{R_l}(\omega,r)\,|^2\big]d\omega\;.
\end{equation}
The Fourier transform with respect to the proper time is
\begin{equation}
\mathcal{G}({\lambda})=\int^{\infty}_{-\infty}e^{i\lambda{\tau}}{G^+}(x,x')d\tau
=\mu^2\sum_{l}\frac{2l+1}{8\pi\lambda}\big[|\,\overrightarrow{R_l}({\lambda}\sqrt{g_{00}},r)\,|^2
+|\,\overleftarrow{R_l}({\lambda}\sqrt{g_{00}},r)\,|^2\big]\theta(\lambda)\;,
\label{gbf}
\end{equation}
where $\theta(\lambda)$ is the step function, and we have used the relation
\begin{equation}
 \sum^l_{m=-l}|\,Y_{lm}(\,\theta,\phi\,)\,|^2= {2l+1 \over
 4\pi}\;.
\end{equation}
The above Fourier transform is hard to evaluate, since we do not know the exact form of the radial functions $\overrightarrow{R_l}(\omega,r)$ and $\overleftarrow{R_l}(\omega,r)$. Here, we choose to compute it both close to the event horizon and at infinity, with the help of the properties of the radial functions in asymptotic regions~\cite{wightman3}:
\begin{equation} \label{asymp1}
\sum_{l=0}^\infty\,(2l+1)\,|\overrightarrow{R}_l(\,\omega,r\,)\,|^2\sim\left\{
                    \begin{aligned}
                 &\frac{4\omega^2}{1-\frac{2M}{r}}\;,\;\;\;\quad\quad\quad\quad\quad\quad\quad r\rightarrow2M\;,\cr
                  &\frac{1}{r^2}
\sum_{l=0}^\infty(2l+1)\,|\,{B}_l\,(\omega)\,|^2\;,\quad\;r\rightarrow\infty
                  \;,
                          \end{aligned} \right.
\end{equation}
\begin{equation} \label{asymp2}
\sum_{l=0}^\infty\,(2l+1)\,|\overleftarrow{R}_l(\,\omega,r\,)\,|^2\sim\left\{
                    \begin{aligned}
                 &\frac{1}{4M^2}\sum_{l=0}^\infty(2l+1)\,|\,{B}_l\,(\omega)\,|^2,\quad\;r\rightarrow2M\;,\cr
                  &\frac{4\omega^2}{1-\frac{2M}{r}},\;\;\;\;\quad\quad\quad\quad\quad\quad\quad\quad r\rightarrow\infty
                  \;.\cr
                          \end{aligned} \right.
\end{equation}
Plugging Eq.~(\ref{gbf}) into Eq~(\ref{abc}), and using
Eq.~(\ref{asymp1}) and ~(\ref{asymp2}), we have
\begin{eqnarray}\label{bab1}
A=B=-C\approx\frac{\gamma_0}{4}[1+f(\omega_0\sqrt{g_{00}},r)]\;,
\end{eqnarray}
in which $f(\omega,r)$ is defined as
\begin{equation}\label{grey}
f(\omega,r)=\frac{1}{4\,r^2\omega^2}\,\sum_{l=0}^\infty\,(2l+1)\,|\,B_l\,(\,\omega)|^2\;,
\end{equation}
and $\gamma_0=\mu^2\omega_0/ 2\pi$ is the spontaneous emission rate.
So, after some straightforward calculations, we obtain, in the
asymptotic regions, \bea \gamma\approx \left\{
  \begin{aligned}
  &-\pi(1-\cos\theta)+\pi^2{\gamma_0\/2\omega_0}
   \sin^2\theta(2+\cos\theta)\big(1+g_{00}\;f(\omega_0\sqrt{g_{00}},2M))\quad\;r\rightarrow2M\;,\cr
  &-\pi(1-\cos\theta)+\pi^2{\gamma_0\/2\omega_0}
\sin^2\theta(2+\cos\theta)\big(1+g_{00}\;f(\omega_0\sqrt{g_{00}},r))\quad\quad\;r\rightarrow\infty\;.\cr
  \end{aligned}
\right.
\eea

The result above shows that, after a single quasi-cycle, the static
atom acquires a geometric phase, which depends on the geometric
properties of its state space. The first term
%$-\pi(1-\cos\theta)$
in the above equation is the one we would have obtained if the
system were isolated from the environment, and the second term is
the correction induced by the interaction between the atom and the
environment. It is interesting to note that, in both  asymptotic
regions, the correction of the geometric phase takes the same form,
which is what one expects  in the Minkowski vacuum\cite{chen,hu2}
multiplied by a factor $(1+g_{00}\;f(\omega_0\sqrt{g_{00}},r))$. The
additional correction as opposed to the Minkowski vacuum, which is
proportional to $g_{00}\;f(\omega_0\sqrt{g_{00}},r)$, can be
attributed to the backscattering of the vacuum field modes off the
space-time curvature of the black hole. This is similar in a sense
to the case of the reflection of the field modes at a reflecting
boundary in a flat space-time. From the definition of the grey-body
factor $f(\omega,r)$ (\ref{grey}) we can see that, in the limit of
$r\rightarrow\infty$, $g_{00}\;f(\omega_0\sqrt{g_{00}},r)$ tends to
be zero, and the result returns to the case in the Minkowski vacuum,
which means that the Boulware vacuum corresponds to our familiar
concept of a vacuum state at large radii.

\subsection{The Unruh vacuum}

Now let us go on our discussion to the Unruh vacuum, which is
supposed to be the vacuum state best approximating the state
following the gravitational collapse of a massive body. The Wightman
function in the Unruh vacuum is given
by~\cite{wightman1,wightman2,wightman3}
 \begin{equation}
{G^+}(x,x')=
\mu^2\sum_{ml}\int^{\infty}_{-\infty}\frac{e^{-i\omega\Delta{t}}}{4
\pi\omega}|\,Y_{lm}(\theta,\phi)\,|^2\bigg[\frac{|\,\overrightarrow{R_l}(\omega,r)\,|^2}{1-e^{-2\pi\omega/\kappa}}
+\theta(\omega)|\,\overleftarrow{R_l}(\omega,r)\,|^2\bigg]d\omega\;,
 \end{equation}
where $\kappa=1/4M$ is the surface gravity of the black hole. Its
Fourier transform is
\begin{eqnarray}\label{guf}
{\cal{G}}(\lambda)&=&\int^{\infty}_{-\infty}e^{i{\lambda}\tau}{G^+}(x,x')d\tau\nonumber\\
&=&\frac{\mu^2}{8\pi{\lambda}}\sum_{l=0}^{\infty}\bigg[\theta({\lambda}\sqrt{g_{00}})(1+2l)|\,\overleftarrow{R_l}({\lambda}\sqrt{g_{00}},r)\,|^2
+\frac{(1+2l)|\,\overrightarrow{R_l}({\lambda}\sqrt{g_{00}},r)\,|^2}{1-e^{-2\pi{\lambda}\sqrt{g_{00}}/\kappa}}\bigg]\;,
\end{eqnarray}
in which $\kappa_r$ is defined as $\kappa/\sqrt{g_{00}}\;$.
Inserting Eq.~(\ref{guf}) into Eq.~(\ref{abc}), one finds that
\begin{eqnarray}
r\rightarrow2M:\left\{ \begin{aligned}
 &A=-C\approx
 \frac{\gamma_0}{4}\;[1+g_{00}\;f(\omega_0\sqrt{g_{00}},2M)
 +\frac{2}{e^{2\pi\omega_0/\kappa_r}-1}]\;,\\
 &B\approx
 \frac{\gamma_0}{4}\;[1+g_{00}\;f(\omega_0\sqrt{g_{00}},2M)]\;,
 \end{aligned} \right.
 \end{eqnarray}
 and
\begin{eqnarray}
r\rightarrow\infty:\left\{ \begin{aligned}
 &A=-C\approx
 \frac{\gamma_0}{4}\;[1+g_{00}\;f(\omega_0\sqrt{g_{00}},r)+\frac{2}{e^{2\pi\omega_0/\kappa_r}-1}
 g_{00}f(\omega_0\sqrt{g_{00}},r)]\;,\\
 &B\approx
 \frac{\gamma_0}{4}\;[1+g_{00}\;f(\omega_0\sqrt{g_{00}},r)]\;.
 \end{aligned} \right.
 \end{eqnarray}
Straightforward calculations lead to \bea\label{uvac}
\gamma\approx\left\{
\begin{aligned}
 &-\pi(1-\cos\theta)+\pi^2{\gamma_0\/2\omega_0}\sin^2\theta\times\\
 &~\bigg[(2+\cos\theta)\big(1+g_{00}\;f(\omega_0\sqrt{g_{00}},2M)\big)
+\frac{2}{e^{2\pi\omega_0/\kappa_r}-1}\cos\theta\bigg]\;,\quad\quad\quad\quad\quad\quad r\rightarrow2M\\
 &-\pi(1-\cos\theta)+\pi^2{\gamma_0\/2\omega_0}\sin^2\theta\times\\
 &~\bigg[(2+\cos\theta)\big(1+g_{00}\;f(\omega_0\sqrt{g_{00}},r)\big)
+g_{00}\;f(\omega_0\sqrt{g_{00}},r)\frac{2}{e^{2\pi\omega_0/\kappa_r}-1}\cos\theta\bigg]\;.~~ r\rightarrow\infty
 \end{aligned} \right.
\eea

Form the equations above, we can see that, in the vicinity of the
event horizon, the first term  in the environment induced geometric
phase is the same as the one in the Boulware vacuum, and the second
one is proportional to a Planckian factor. At infinity, the
Planckian factor is modified by a grey-body factor
$g_{00}\;f(\omega_0\sqrt{g_{00}},r)$, which is caused by the
backscattering off the space-time curvature. In order to have a
better understanding of the features of the geometric phase  and of
what it tells us about, it would be rewarding to examine here the
geometric phase of a static atom in a thermal bath at temperature
$1/\beta$ in a flat space-time. In this  case, the Wightman function
reads
\begin{equation}
{G^+}(x,x')=-{1\/4\pi^2}\sum_{m=-\infty}^{\infty}{1\/(t-t'-im\beta-i\varepsilon)^2
-|{\bf x-x^\prime}|^2}\;,
\end{equation}
and similar calculations leads to
 \beq\label{ther}
\gamma\approx-\pi(1-\cos\theta)+\pi^2{\gamma_0\/2\omega_0}\sin^2\theta
 \bigg(2+\cos\theta+\frac{2}{e^{\beta\omega_0}-1}\cos\theta\bigg)\;.
\eeq Comparing Eq.~(\ref{uvac}) and Eq.~(\ref{ther}), one finds
that, apart from the backscattering of the space-time curvature, the
result is the same as that in the case of a two-atom system immersed
in a thermal bath at an effective temperature
$T=\kappa_r/2\pi=T_H/\sqrt{g_{00}}$ in a flat space-time, where
$T_H=\kappa/2\pi$ is the Hawking temperature~\cite{hk}. This clearly
suggests that at the horizon, there is a thermal flux going
outwards. However,
 when
$r\rightarrow 2M$, the temperature $T$ is divergent, since this
effective temperature is a joint effect of both the thermal flux
from the black hole and the Unruh effect due to that the system is
accelerating with respect to the local free-falling inertial frame
so as to maintain at a fixed distance from the black hole, and the
acceleration diverges at the horizon. As the atoms are placed
farther away from the black hole, the thermal radiation becomes
weaker due to the backscattering off the space-time curvature. At
the infinity, the grey-body factor vanishes and it returns to the
Minkowski vacuum case. This suggests that, in the Unruh vacuum, no
thermal radiation is felt at the infinity due to the backscattering
of the outgoing thermal radiation off the space-time curvature.

\subsection{The Hartle-Hawking vacuum}

Finally, let us come to the Hartle-Hawking vacuum. The Wightman function in this vacuum is given by \cite{wightman1,wightman2,wightman3}
\begin{equation}
{G^+}(x,x')=\mu^2\sum_{ml}\int^{\infty}_{-\infty}\frac{|\,Y_{lm}(\theta,\phi)\,|^2}{4
\pi\omega}\bigg[\frac{e^{-i\omega{\Delta{t}}}}{1-e^{-2\pi\omega/\kappa}}|\,\overrightarrow{R_l}(\omega,r)\,|^2+\frac{
e^{i\omega\Delta{t}}}{e^{2\pi\omega/\kappa}-1}|\,\overleftarrow{R_l}(\omega,r)\,|^2\bigg]d\omega\;,\nonumber\\
\end{equation}
and its Fourier transform is
\begin{equation}
{\cal{G}}(\lambda)=\int^{\infty}_{-\infty}e^{i{\lambda}\tau}{G}^+(x,x')d\tau
=\mu^2\sum_{l=0}^{\infty}\frac{(1+2l)}{8\pi{{\lambda}}}\bigg[
\frac{|\,\overrightarrow{R_l}({\lambda}\sqrt{g_{00}},r)\,|^2}{1-e^{-2\pi{{\lambda}}/\kappa_r}}+
\frac{|\,\overleftarrow{R_l}(-{\lambda}\sqrt{g_{00}},r)\,|^2}{1-e^{-2\pi{{\lambda}}/\kappa_r}}\bigg]\;.
\end{equation}
Similar calculations then lead to
\begin{eqnarray}
r\rightarrow2M:\left\{ \begin{aligned}
 &A=-C\approx
 \frac{\gamma_0}{4}\frac{e^{2\pi\omega_0/\kappa_r}+1}{e^{2\pi\omega_0/\kappa_r}-1}\;
 [1+g_{00}\;f(\omega_0\sqrt{g_{00}},2M)]\;,\\
 &B\approx
 \frac{\gamma_0}{4}\;[1+g_{00}\;f(\omega_0\sqrt{g_{00}},2M)]\;,
 \end{aligned} \right.
 \end{eqnarray}
\begin{eqnarray}
r\rightarrow\infty:\left\{ \begin{aligned}
 &A=-C\approx
 \frac{\gamma_0}{4}\frac{e^{2\pi\omega_0/\kappa_r}+1}{e^{2\pi\omega_0/\kappa_r}-1}\;
 [1+g_{00}\;f(\omega_0\sqrt{g_{00}},r)]\;,\\
 &B\approx
 \frac{\gamma_0}{4}\;[1+g_{00}\;f(\omega_0\sqrt{g_{00}},r)]\;,
 \end{aligned} \right.
 \end{eqnarray}
and
\bea
\gamma\approx\left\{\begin{aligned}
&-\pi(1-\cos\theta)+\pi^2{\gamma_0\/2\omega_0}
\sin^2\theta\times\\
&\bigg[(2+\cos\theta)\big(1+g_{00}\;f(\omega_0\sqrt{g_{00}},2M)\big)
+\frac{2}{e^{2\pi\omega_0/\kappa_r}-1}\big(1+f(\omega_0\sqrt{g_{00}},2M)\big)\cos\theta\bigg]\;,\\ &\qquad\qquad\qquad\qquad\qquad\qquad\qquad\qquad\qquad\qquad\qquad\qquad\qquad\qquad\qquad r\rightarrow2M\\
&-\pi(1-\cos\theta)+\pi^2{\gamma_0\/2\omega_0}
\sin^2\theta\times\\
&\bigg[(2+\cos\theta)\big(1+g_{00}\;f(\omega_0\sqrt{g_{00}},r)\big)
+\frac{2}{e^{2\pi\omega_0/\kappa_r}-1}\big(1+f(\omega_0\sqrt{g_{00}},r)\big)\cos\theta\bigg]\;.\\
&\qquad\qquad\qquad\qquad\qquad\qquad\qquad\qquad\qquad\qquad\qquad\qquad\qquad\qquad\qquad r\rightarrow\infty
\end{aligned} \right.
\eea

Here, unlike that in the Unruh vacuum case, the geometric phase
takes the same form in two asymptotic regions.  The first term is
the same as that in the Boulware vacuum, and the second term is
composed by two parts, one is proportional to the standard Planckian
factor and the other is proportional to a Planckian factor modified
by a grey-body factor caused by the backscattering off the
space-time curvature. This suggests that there are thermal radiation
outgoing from the horizon and that incoming from infinity, both of
which are weakened by the backscattering off the curvature on their
way. In both  asymptotic regions, the impact of the black hole to
the geometric phase is the same as that of a thermal bath at
temperature $T=T_H/\sqrt{g_{00}}$ in a flat space-time. As
$r\rightarrow \infty$, the effective temperature becomes the Hawking
temperature, since the acceleration needed to maintain the two-level
atom system at a fixed radial distance vanishes, and the temperature
is purely due to the thermal bath the black hole is immersed in.
Therefore, the Hartle-Hawking vacuum does not correspond to our
familiar concept of vacuum. It is actually a state that describes a
black hole in equilibrium with an infinite sea of blackbody
radiation~\cite{wightman3}.

\section{conclusion}
 We have studied the evolution of a two-level atom
interacting with a bath of fluctuating quantum scalar fields at a
fixed radial distance outside a Schwarzschild black hole in the
framework of open quantum systems, and calculated the geometric
phase of the atom in the Boulware, Unruh and Hartle-Hawking vacua
respectively. We find, for all the three cases, that the geometric
phase of the atom turns out to be affected by space-time curvature
which backscatters the vacuum field modes. In the Unruh vacuum, the
geometric phase behaves as if there were an outgoing flux of thermal
radiation which is backscattered by the space-time curvature. In the
Hartle-Hawking vacuum, the geometric phase exhibits behaviors as if
the atom were in a thermal bath at an effective temperature which
reduces to the Hawking temperature at infinity.

Generally speaking, the geometric phase is of significance only if we compare the situation of the atom to some other situations. For our current case, it is natural to compare the atom kept at a fixed position to the one released free. However, the calculation of the geometric phase of a free-falling atom, although of great physical importance, is  technically  much more difficult  than the situation considered in our current paper. The reason is that in a Schwarzschild space-time, the vacuum fluctuations, namely, the environment the atom is coupled to, varies with the radial distance from the black hole. We will leave this to a future investigation. Nevertheless, as it has already been done in a recent experiment~\cite{critical}, we may manage to measure the decoherence factor of the off-diagonal elements of the reduced density matrix directly in order to get the geometric phase. So, we can obtain the geometric phase in a Schwarzschild space-time and that in a flat space-time respectively, and then compare the results in the two situations to reveal the effects of Hawking radiation.  In summary, our analysis shows, at least in principle, that geometric phase might provide a potentially new way to detect the Hawking radiation, albeit  difficulties of constructing specific experimental procedures that can actually be carried out.

\begin{acknowledgments}
 This work was supported in part by the National Natural Science Foundation of China under
Grants No. 11075083, and No. 10935013; the Zhejiang Provincial
Natural Science Foundation of China under Grant No. Z6100077; the
National Basic Research Program of China under Grant No.
2010CB832803; the Program for Changjiang Scholars and Innovative Research Team in University under Grant No. IRT0964; and the Hunan
Provincial Natural Science Foundation of China under Grant No.
11JJ7001.
\end{acknowledgments}

\end{document}